# Quantum transport properties of *β*-Bi$_4$I$_4$ near and well beyond the extreme quantum limit

Peipei Wang[1,2], Fangdong Tang[2], Peng Wang[2], Haipeng Zhu[3], Chang-Woo Cho[2], Junfeng Wang[3], Xu Du[4], Yonghong Shao[1*], Liyuan Zhang[2*]

1. Key Laboratory of Optoelectronic Devices and Systems of Ministry of Education and Guangdong Province, College of Optoelectric Engineering, Shenzhen University, 518060 Shenzhen, China.
2. Department of Physics, Southern University of Science and Technology, 518055 Shenzhen, China.
3. Wuhan National High Magnetic Field Center and School of Physics, Huazhong University of Science and Technology, Wuhan 430074, China.
4. Department of Physics and Astronomy, Stony Brook University, Stony Brook NY 11794, US

## Abstract

We have investigated the magneto-transport properties of *β*-Bi$_4$I$_4$ bulk crystal, which was recently theoretically proposed and experimentally demonstrated to be a topological insulator. At low temperature ***T*** and magnetic field ***B***, a series of Shubnikov-De Haas(SdH) oscillations are observed on the magnetoresistivity (MR). The detailed analysis reveals a light cyclotron mass of 0.1 $m_e$, and the field angle dependence of MR reveals that the SdH oscillations originate from a convex Fermi surface. In the extreme quantum limit (EQL) region, there is a metal-insulator transition occurring soon after the EQL. We perform the scaling analysis, and all the isotherms fall onto a universal scaling with a fitted critical exponent $\zeta \approx 6.5$. The enormous value of critical exponent $\zeta$ implies this insulating quantum phase originated from strong electron-electron interactions in high fields. However, in the far end of EQL, both the longitudinal and Hall resistivity increase exponentially with ***B***, and the temperature dependence of the MR reveals an energy gap induced by the high magnetic field, signifying a magnetic freeze-out effect. Our findings indicate that bulk *β*-Bi$_4$I$_4$ is an excellent candidate for a 3D topological system for exploring EQL physics and relevant exotic quantum phases.

* corresponding author: shaoyh@szu.edu.cn, zhangly@sustech.edu.cn

# Introduction

Topological insulators (TIs) are quantum materials that have insulating bulk states and metallic surface states consisting of massless Dirac fermions[1-6]. The surface states are robust against time-reversal invariant perturbations and could lead to many interesting transport phenomena such as weak antilocalization[7-9] [1–3], quantum oscillations[10-12], and Aharonov-Bohm effect[13-15]. By far, Bismuth-based binary or ternary compounds such as 3D $Bi_{1-x}Sb_x$, quasi-2D $Bi_2Se_3$, $Bi_2Te_3$, $Bi_2Te_2Se$ have become the essential class of TIs. These TIs had been studied extensively in the past years and continues to attract research interests, due to promising both rich novel physics and potential applications in spintronics and quantum computation.

Recently, a new topological insulator $β$-$Bi_4I_4$ has been theoretically predicted and experimentally confirmed[16-18]. However, ARPES experiments give different results about the topological classification of $β$-$Bi_4I_4$. Gabriel Autès *et. al* had performed the measurements on the (001) surface and revealed a single Dirac cone located at the M point of the surface Brillouin zone[17], confirmed $β$-$Bi_4I_4$ as a strong TI. But in a recent work, Ryo Noguchi *et. al* reported that only the side surface (the (100) plane) hosts Dirac surface state, indicate that $β$-$Bi_4I_4$ is a weak TI[18]. Despite the ambiguity in its classification, $β$-$Bi_4I_4$ has shown many fascinating transport properties. Hysteretic profile forms at around 300K on the resistivity-temperature curve, indicating a structural phase transition between $β$-$Bi_4I_4$ and $α$-$Bi_4I_4$[18]. The theoretical calculations and the experiments found a pressure-induced superconducting phase and multi-structural or topological phase transitions in $β$-$Bi_4I_4$[19-22]. The first-principles calculations found that $β$-$Bi_4I_4$ has a very similar bulk electronic structure to that of $α$-$Bi_4I_4$, with a 3D Fermi surface in the valence band and a 2D Fermi surface in the conduction band[23].

In this work, we carried detailed studies on the angular and temperature dependence of magneto-transport properties of $β$-$Bi_4I_4$ with the steady and extra high pulsed magnetic field. Through SdH oscillations, we obtained a small cyclotron mass of 0.1 $m_e$, and a 3D angle mapping for magneto-resistivity (MR) reveals a convex Fermi surface. With the entrance of EQL, there occurs a metal-insulator transition. In the region far away from the EQL, the MR increases exponentially regard to ***B***, and a field induced effective gap is observed. Our findings indicate that $β$-$Bi_4I_4$ is possible an ideal platform for exploring quantum limit physics and relevant exotic phases.

# Experimental setup

Single crystals of $\beta$-Bi$_4$I$_4$ were grown using the methods described in Ref[17]. Starting materials Bi and HgI$_2$ powder with mole ratio 1:2 were mixed and finely grounded in a glove box, then loaded in a quartz tube 300 mm long and 16 mm in diameter. The quartz tube was flame sealed under high dynamic vacuum and placed into a two-zone furnace. A temperature gradient of 250°C - 210°C was applied. After 20-30 days, the quartz tube was taken out and cooled in the air. Needle-like crystals with typical size 5 mm×0.3 mm×0.2 mm were obtained. The crystal structures and face index were examined by powder X-ray diffraction recorded by Rigaku smartlab 9 KW with Cu $K_\alpha$ radiation ($\lambda$ = 1.5406 °A), average stoichiometry was determined by energy-dispersive x-ray spectroscopy. The transport properties under magnetic field were carried out in Oxford TeslatronPT with temperature variable from 1.5 K to 300 K and magnetic field up to 14 Tesla. The 52 T pulsed magnetic field measurements were performed at Wuhan National High Magnetic Field Center. Before making the electrical connections, the crystal surface was cleaned by Argon plasma followed by a deposition of 5/30 nm thick Cr/Au Hall bar pattern through a mask. Both the surface cleaning and Cr/Au deposition are crucial to the measurements, and it ensures the Ohmic contacts with a typical contact resistance of ~ 1 Ohm. By contrast, directly use silver paste resulted in large contact resistance of 0.2~2 KOhm, which increases further with time. This is possible because the surfaces of $\beta$-Bi$_4$I$_4$ are highly susceptible to influence by moisture and organic solvent.

## Results and discussion

$\beta$-Bi$_4$I$_4$ has a quasi-1D structure, see Fig. 1(a), with Bi-I molecular chains stack along both $a$ and $c$ axis through weak noncovalent interactions. The chain's direction is along axis $b$, as indicated in Fig. 1b inset. The average atomic ratio determined from EDX is Bi:I = 0.51 : 0.49, and no traces of mercury were detected on the cleaved surfaces. The structure of $\beta$-Bi$_4$I$_4$ is very similar to that of $\alpha$-Bi$_4$I$_4$. One crucial feature which distinguishes the $\alpha$- phase is the presence of (005) and (007) reflection peaks located at $2\theta \approx 22.3°$ and $2\theta \approx 31.5°$ [18,23]. Fig. 1(b) shows that none of the two peaks is observed, confirming that our crystals are $\beta$-Bi$_4$I$_4$. Another feature that distinguishes these two phases is the temperature-dependent resistivity, with the $\alpha$- phase being much more resistive[18]. Fig. 1(c) presents a typical curve of resistivity $\rho$ as a function of temperature $T$ for our $\beta$-Bi$_4$I$_4$ crystals. As $T$ decreases from room temperature, the resistivity of the freshly prepared samples (black line) usually forms a wide peak centered at around 200~220K, followed by a metallic behavior down to 50K, then increases again till the lowest temperature. As a comparison, we also present a ρ(T) curve of another sample in the inset, clear hysteretic loop at 300K, and the much more resistive behavior below 200K indicates the formation of α-Bi4I4 after the phase transition at 300K. We also found that resistivity is readily influenced by the environment. After warming up to 300K and exposed to the air for a period, the ρ(T) changes with the peak shifts to lower temperature and the resistivity at low $T$ increases. The total 10 hours of exposure (blue line) resulted in a semiconductor behavior quite similar to the α-phase Bi4I4. We speculate that the reaction of iodine with moisture may change the surface conductivity.

We further character the sample with magneto-transport measurements. As shown in Fig. 1(d), when the magnetic field is applied along the z direction, the longitudinal $\rho_{xx}$ shows clear SdH oscillations. At least three peaks can be distinguished, corresponding to the Landau level index $N$ = 3, 2, and 1. Another prominent feature is that the $\rho_{xx}$ decrease sharply above **B** = 10.4T, to a negative- MR region. A simple FFT analysis yields a single frequency, indicating that only one pocket is on the Fermi surface. The single-band nature is further confirmed by the Hall resistivity $\rho_{xy}$, which shows a linear behavior from -13.5 T to 13.5 T. The bulk carrier density $n$ and the mobility $\mu$ at 1.5 K are obtained from the linear fitting of $\rho_{xy} = $ **B**$/ne$ and via $\sigma_0 = ne\mu$, which yield $n$ = 1.2 × $10^{18}$/cm$^3$ (hole type) and $\mu$ = 5400 cm$^2$V$^{-1}$s$^{-1}$.

To understand the nature of the SdH oscillations, we performed magnetic field angle-dependent measurements to tracked how the SdH oscillations shift with $\theta$ (the tilted angle between the magnetic field and the -$a$ axis). Fig. 2(a) presents the $\rho_{xx}$ at different angles $\theta$, here all the curves have been shifted for clarity. As $\theta$ increases from 0 (where **B** // -$a$), the fields of $N$ =1 and $N$ =2 peaks decrease until $\theta = 67.5°$, then began to increase. The minimal angle 67.5° is very close to 72° between the (001) and (100) planes of β-Bi4I4. After an FFT analysis, the corresponding frequency as a function of $\theta$ is shown in Fig. 2(b). According to the Onsager relation $F = S_F \hbar /2\pi e$, the oscillation frequency is proportional to the extremal cross-section area of the Fermi surface $S_F$. For a two-dimensional (2D) Fermi surface, the oscillation frequency depend only on the field component **B** = **B**$_0$cos$\theta$. As $\theta$ increase, the frequency will increase the following relation of $F(\theta) = F_0/\cos\theta$. Choose the minimal frequency as $F_0$ when $\theta = 67.5°$ (which is 9.2T), we have plotted the $F_0/\cos(\theta - 67.5°)$ in Fig.2(b), as the red line shows. The measured data fit with the cosine relation only when $15° \leq \theta \leq 67.5°$, indicate that the oscillations originate from a 3D Fermi surface rather than a 2D, but with a convex shape. This is consistent with the first-principles calculation[23]. Fixing the angle at 67.5°, if we assume a circular Fermi surface cross-section, the $k_F$ can be calculated from $S_F = \pi k_F^2$, which is $k_F = 1.7 \times 10^{-2}$ Å$^{-1}$. Fig. 2(b) inset shows the interceptions obtained from the Landau level fan diagram, from which the Berry's phase can be determined. In 3D system, an interception in the range of $\pm 1/8$ implies a nontrivial Berry's phase of π, and it is trivial otherwise. As the magnetic field tilted from **B**//-a, the Berry's phase becomes nontrivial when $\theta \geq 15°$, in accordance with the ARPES results in ref[18], where the surface Dirac cone is found at the side surface ((100) surface) of the crystal.

After subtracting the background of $R_{xx}$ or $\rho_{xx}$ by taking a second derivative, the SdH amplitudes taken at different temperatures as a function of reciprocal **B** are presented in Fig. 2(c), along with the FFT spectrum. The cyclotron mass of carrier can be obtained by fitting the temperature dependence of the SdH amplitudes following Lifshitz-Kosevich (LK) theory:

$$A(T) = A_0 \frac{2\pi^2 k_B T/\hbar\omega_c}{\sinh(2\pi^2 k_B T/\hbar\omega_c)}$$

Where $k_B$ is the Boltzmann's constant, and $\omega_c$ is the cyclotron frequency $\omega_c = e\boldsymbol{B}/m^*$. The effective cyclotron mass $m^*$ is determined to be 0.1 $m_0$ (where $m_0$ is the free electron mass) in Fig. 2(d).

Such a small effective mass implies large cyclotron energy $\hbar\omega_c$ in magnetic fields. When the field **B** is high enough that the criterion $\hbar\omega_c \gg E_F \gg k_B T$ is satisfied, the system enters the EQL where all the electrons occupy only the lowest Landau level. Experimentally, the EQL can be identified by the last peak of quantum oscillations on resistivity or thermoelectric curves[24,25]. In Fig. 1(d), $\rho_{xx}$ decreases sharply after the $N$ = 1 peak at 10T, suggesting that the system approaches the EQL. To investigate the quantum phases of β-Bi$_4$I$_4$ beyond the EQL, we further carried out the pulsed high magnetic field transport measurements up to 43T, where the direction of magnetic field is pointed around the c axis, as shown in Fig. 3(a) and 3(c). The quantum oscillations end up at 13.4T with the $\rho_{xx}$ reaches the minima. Above 13.4T, the $\rho_{xx}$ began to increase monotonically without saturate, meanwhile the Hall resistivity $\rho_{xy}$ changes its behavior from temperature independent to temperature-dependent. We believe that both the sign change in $\rho_{xx}$ from negative to positive and the slope change in $\rho_{xy}$ indicate the system to enter the EQL at 13.4T, where the Fermi energy crosses only the lowest Landau level band $N$ = 0 and lies right below the bottom of the band $N$ = 1[26].

As the magnetic field increases further, there exists a critical field of $B_c$ = 15.9 T where all the $\rho_{xx}$ curves taken at different temperatures share a common crossing point (Fig. 3(a) inset). We also plot $\rho_{xx}$ as a function of temperature $T$ with fixed $B$ ranging from 13T to 20T, as shown in Fig. 3(b). Below the critical field $B_c$, the $\rho_{xx}$ decreases monotonically as T decreases, while increases when the field is above $B_c$. This profile signals a metal-insulator transition at the critical field: the system is metallic below $B_c$ and insulating above $B_c$. Typically, for magnetic field induced metal-insulator transition, the scaling relation of the resistivity to temperature and magnetic field can be described by[27,28]

$$\rho(B,T) = \rho(B_c)f\left[|B - B_c|T^{-1/\zeta}\right]$$

Where $f(x)$ is a scaling function with $f(0) = 1$, and $\zeta$ is the critical exponent which can be determined by evaluating the inverse slope of the log-log plot of $\left.\frac{d\rho_{xx}}{dB}\right|_{B_c}$ vs. $1/T$[29]. With $B_c$ and $\zeta$ in hand, the scaling function can be directly tested without applying other fitting parameter. In Fig. 3(d), we performed such a scaling analysis, and indeed all the isotherms fall onto a nearly single curve as a function of $|B - B_c|T^{-1/\zeta}$. In Fig. 3(d) inset, we obtained $1/\zeta$ = 0.153 ± 0.03 from the linear fitting of the data, this yields the critical exponent of $\zeta \approx 6.5$. According to a recent theory which describes the field induced metal-insulator transition in 3D systems[30], a large critical exponent $\zeta \approx 6$ implies that strong electron-electron interactions may involved at high fields.

In Fig. 4(a), however, with a much higher magnetic field $B$ (>25T), the $\rho_{xx}$ grows exponentially with increasing $B$. We have ploted the logarithm resistivity $\ln(\rho_{xx})$ as a function of $B$ under various temperatures (the curves have been shifted for clarity). It can be clearly seen that $\ln(\rho_{xx})$ increase linearly with $B$ from 20~25T to 43T, depending on the sample temperature. This exponential increasing $\rho_{xx}$ is quite different from those $B^2$, $B$, or $\sqrt{B}$ relations as usually seen in topological semimetals under high fields[31]. This new insulator region can be attributed to the magnetic freeze-out effect that led to a reduced carrier concentration. Typically, the magnetic freeze-out arises from the electron localization in the extrinsic hopping conductivity region[32,33]. At low temperatures, the electrons are localized at impurity or dopant ions, and the conduction mechanism in these materials is variable range hopping between the impurity or dopant. The high magnetic field will further compress the wave functions of the localized states and reduce the overlap of the electron orbitals between neighbor ions, resulting in an exponential MR. Another mechanism for the magnetic freeze-out effect is an increasing effective energy gap with the field in an intrinsic narrow band-gap semiconductor with small carrier masses[34]. Under such conditions, at high fields where $\hbar\omega_c \gg k_B T$ holds, the carrier concentration decreases exponentially with magnetic field, $\frac{n(B)}{n(0)} \propto \exp(-\hbar\omega_c/4k_B T)$, which results in the exponentially increase of both $\rho_{xx}$ and $\rho_{xy}$ with field $B$. Although both mechanisms exhibit an exponentially increasing magneto-resistivity, they are distinguishable in transport measurements. The distinctive evidence is that the argument of the exponential for the former is quadratic[35,36], while it is linear for the latter one[34]. In Fig. 4(b) we plot the $\ln(\rho_{xx})$ as a function of inverse temperature $1/T$ under fixed field $B$, then fit the high temperature data to extract the effective band gap following Arrhenius plotting, $\rho_{xx} \propto \exp(\frac{\Delta}{2k_B T})$. The obtained band gap with error bar are shown in Fig. 4(c), it increase almost linearly with magnetic field from zero at 15.9T to 12K at 43T. As a result, the concentration of the thermal carriers decreases which leads to the exponential behavior as shown in Fig. 4(a). This is further confirmed by the Hall resistivity (Fig. 4(d)), where the magnetic field up to 52T is parallel to the z axis. A clear non-linear profile can be seen, and the high field data ($>$ 25T) is proportional to $e^B$. Our data support the magnetic freeze-out effect of carriers concentration mechanism well beyond the EQL. However, as the impurities and defects are inherently presented in crystals, both intrinsic and extrinsic mechanism could be involved simultaneously in a particular sample. Further investigations are needed for better understanding in the near region of EQL.

## Conclusion

In summary, we have investigated the magneto-transport properties of a small band gap topological insulator $\beta$-Bi$_4$I$_4$ in a magnetic field up to 43T. There exist a critical field $B_c = 15.9$T where metal to insulator transition happens. Below $B_c$ the MR shows clear SdH oscillations, which reveals a 3D convex Fermi surface and a small effective hole mass of $0.1m_e$. In the near region of EQL, there is a metal-insulator transition around $B_c$ Furthermore, the result of scaling analysis suggests resulting in a strong electron-electron transition. Far away from the EQL, both the longitudinal and Hall resistivity increase exponentially with field $B$. The temperature dependence of the MR reveals an energy gap that increases linearly with field $B$, which signifies the magnetic freeze-out of carrier concentration, where the system is with a narrow bulk energy gap. The extract mechanisms of the transitions observed here still require further investigation.

## Acknowledgements

We thank Haizhou Lu, Kun Yang, and Haiwen Liu for their enlightening discussions. The Work was supported by Guangdong Innovative and Entrepreneurial Research Team Program (No. 2016ZT06D348), NFSC (11874193), China postdoctoral Science Foundation (2020M672760), and Shenzhen Fundamental subject research Program (JCYJ20170817110751776). X.D. acknowledges support from NSF under award DMR-1808491.

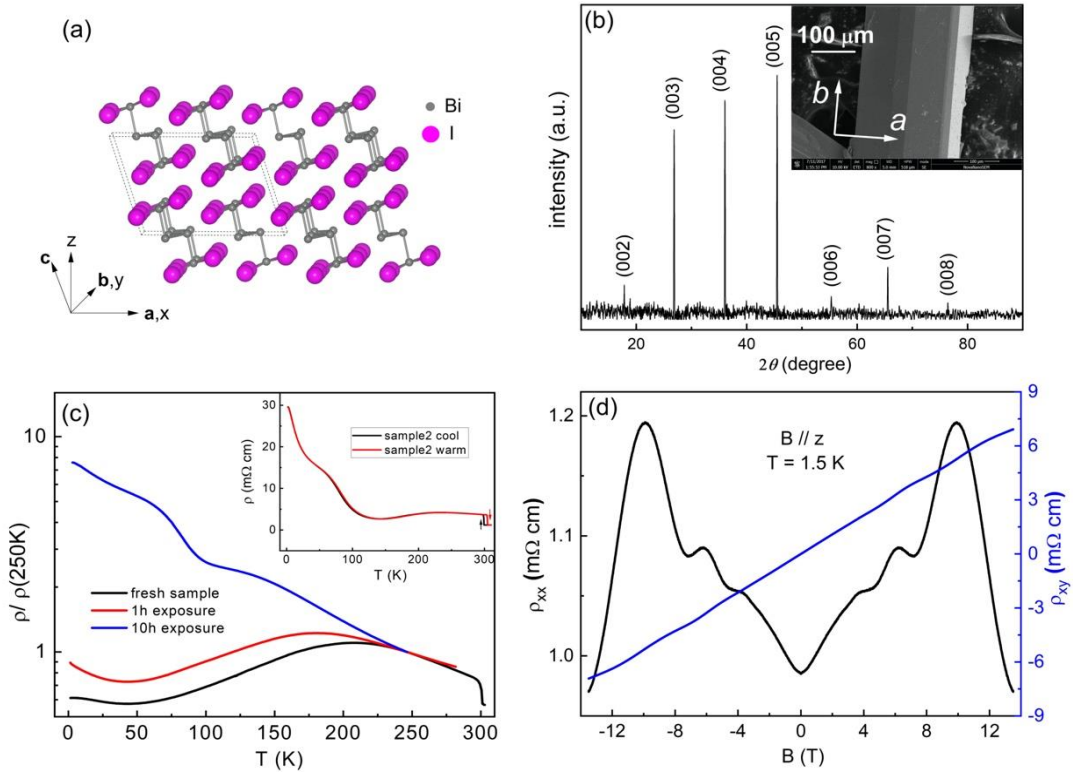

**Figure 1.** (a) The $\beta$-Bi$_4$I$_4$ crystalline in space group $C12/m1$(No.12), with lattice parameters $a$ = 14.353 Å, $b$ = 4.423 Å, $c$ = 10.476 Å, and $\beta$ = 107.8° between the $a$ and $c$ axis. Pink and grey spheres represent the Bismuth and iodine ions, respectively, and the grey dashed box draws the unit cell. (b) The XRD pattern of $\beta$-Bi$_4$I$_4$ single crystals with (00$l$) reflections. Inset: A SEM image of a ribbon-shaped sample usually has quite irregular facets. (c) Temperature dependence of $\rho_{xx}$ curves normalized to $\rho(250K)$ for sample 1 with different exposure times in the air. The fresh sample refers to the sample with the least exposure time, which was measured immediately after the contacts were made. Upper inset: RT curves of sample 2 show an insulating behavior of $\alpha$ phase. A hysteresis profile can be seen at 300K. (d) $\rho_{xx}$ and $\rho_{xy}$ of sample 1 taken at 1.5K, the magnetic field is parallel to the $z$ direction.

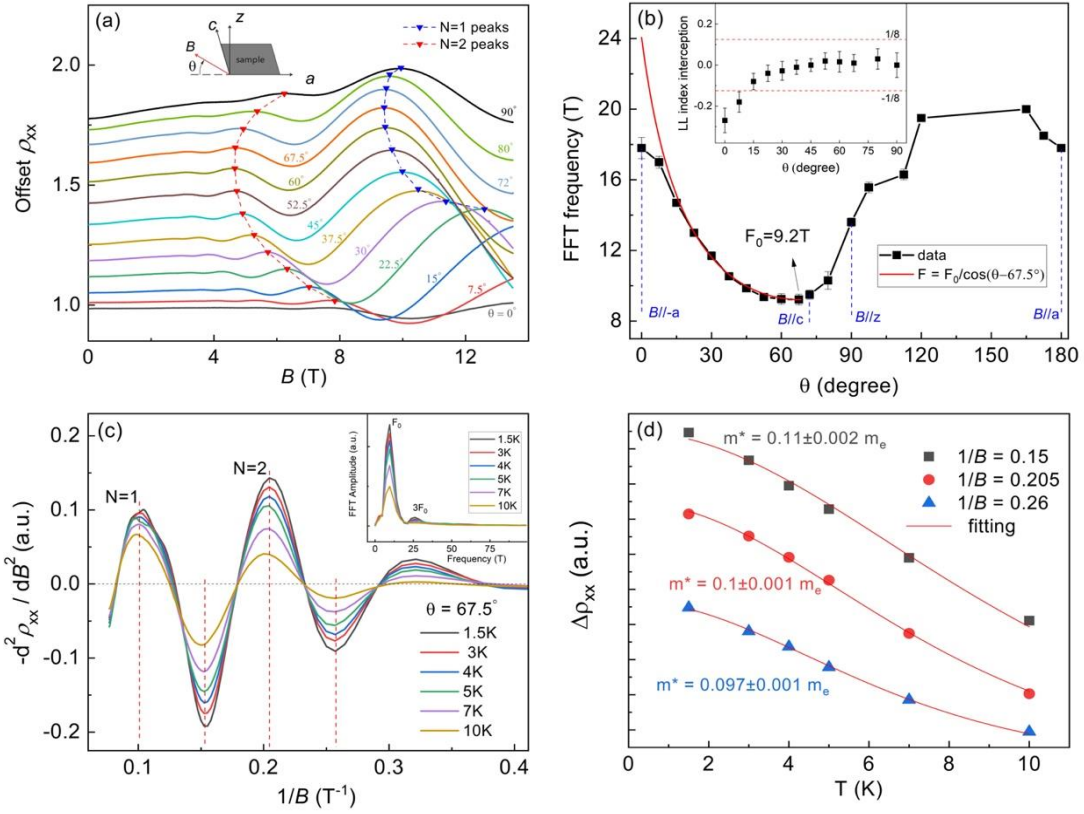

**Figure 2**. (a) The angle dependence of the SdH oscillations from $\theta = 0$ to $\theta = 90°$, the blue and red triangle marks the $N = 1$ and $N = 2$ Landau level peaks respectively. (b) The obtained oscillation frequency as a function of $\theta$ from 0 to 180°. The red line draws the angle dependence of the frequency for a 2D Fermi surface, $F_0$ corresponds to the minimal cross-section area at $\theta = 67.5°$. The inset shows the interception extracted from the Landau level (LL) fan diagram. (c) $-d^2\rho_{xx}/d\boldsymbol{B}^2$ as a function of $1/\boldsymbol{B}$, at different temperatures range between 1.5K and 10K. Inset shows the FFT transform of the data in the main panel. (d) The Normalized oscillation amplitudes as a function of temperature, taken in different fields. The red line shows the fitting of the effective cyclotron mass following Lifshitz-Kosevich theory.

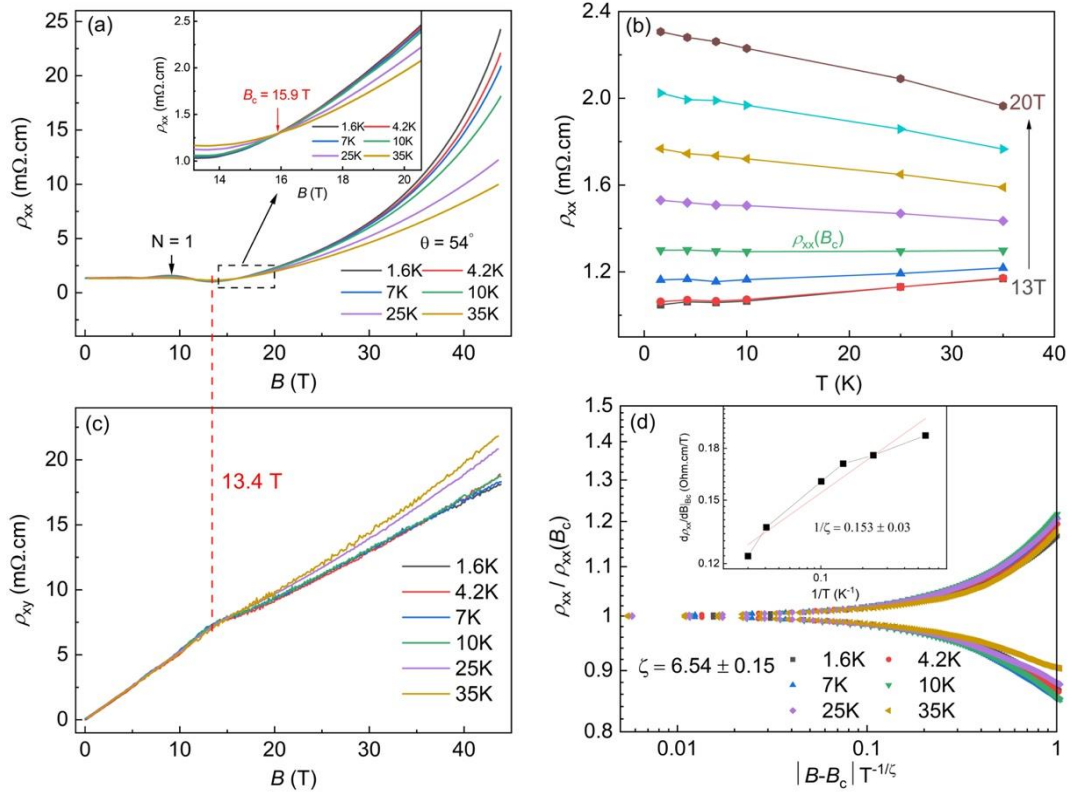

**Figure 3.** (a) The field dependence of $\rho_{xx}$ up to 43T under different temperatures. The red dashed line marks the ending of the quantum oscillation, above which the system enters the EQL. Inset shows the metal to insulator transition happened at a critical field of 15.9T. (b) Plot the $\rho_{xx}$ as a function of temperature at fixed magnetic field ***B***. (c) The field dependence of $\rho_{xy}$ up to 43T. (d) Scaling plot of the normalized $\rho_{xx}$, all the isotherms fall onto a nearly single curve as a function of $|B-B_c|T^{-1/\zeta}$ with $\zeta \approx 6.5$, inset shows the linear fitting of the $\log(d\rho_{xx}/d\textbf{\textit{B}})$ at $B_c$ vs. $\log(1/T)$, which yields the $1/\zeta$.

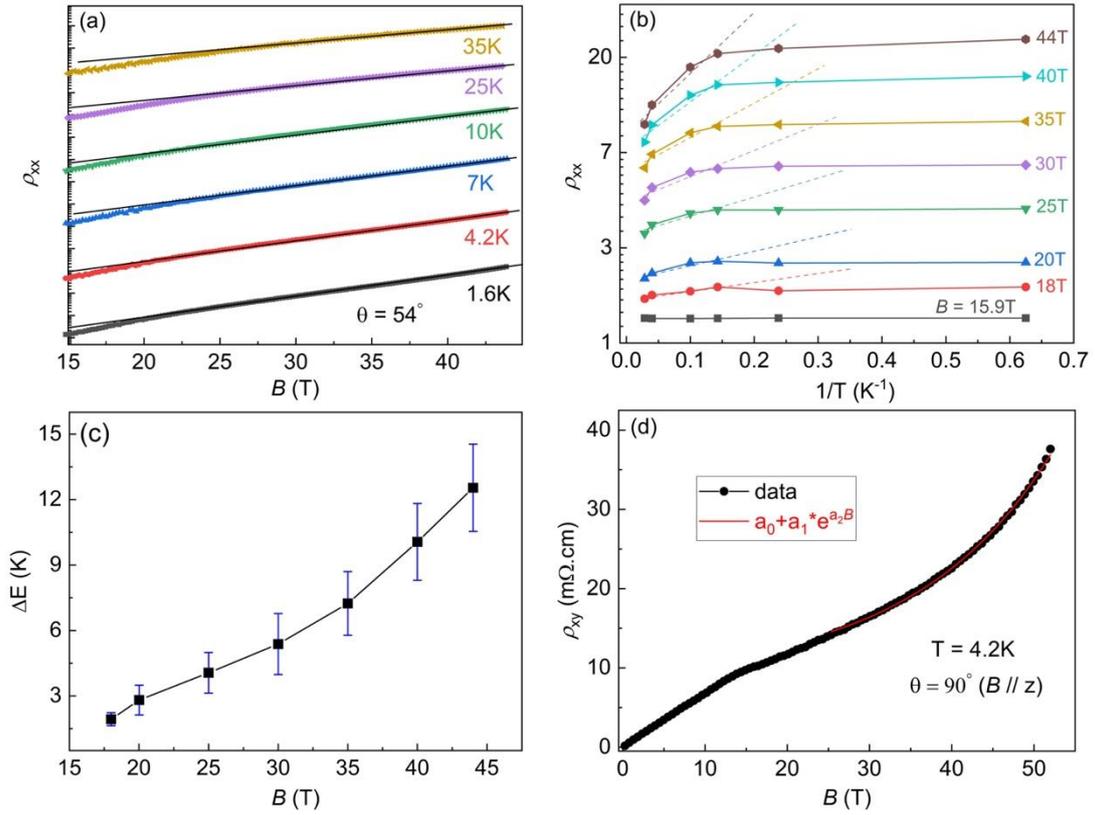

**Figure 4**.  (a)  $\ln(\rho_{xx})$ as a function of field ***B*** up to 43T, the curves have been shifted for clarity, the linear behavior above 20T demonstrates the exponential increase of $\rho_{xx}$. (b) The Arrhenius plot of the temperature dependence of $\rho_{xx}$ under fixed magnetic field range from 15.9T to 43T.  (c) The energy gap is relative to a function of ***B***, and it increases almost linearly with the field.  (d) The Hall resistivity up to 52T at 4.2K, where the field is parallel to the z direction. Under high field, $\rho_{xx}$ also increase exponentially with ***B***, the red line draws a simple $e^B$ fitting.